\documentclass[a4paper,11pt]{article}
\usepackage[toc,page]{appendix}
\usepackage{graphicx}
\usepackage{colortbl}
\date{}
\usepackage{amsmath}
\usepackage{subfigure}
\usepackage{mathtools}
\usepackage{float}
\usepackage{epstopdf}
\usepackage{amsfonts,amsmath,amssymb}
\usepackage{hyperref}

\usepackage{epsfig,multicol,bbm}

\newcommand\fverb{\setbox\pippobox=\hbox\bgroup\verb}
\newcommand\fverbit{\egroup\item[\fbox{\unhbox\pippobox}]}

\newbox\pippobox
\begin{document}
\title{\bf Analytically Approximation Solution to $R^{2}$ Gravity}
\author{\textbf{S. N. Sajadi$^{a,b}$}\thanks{naseh.sajadi@gmail.com}, 
\textbf{Ali Hajilou$^{c,d}$}\thanks{hajilou@mi-ras.ru}, \textbf{S. H. Hendi$^{a,e}$}\thanks{hendi@shirazu.ac.ir}
 \\\\
%\textit{{\small $^a$ Department of Physics, School of Science, Shiraz University, Shiraz 71454, Iran}}\\
\textit{{\small $^a$ Department of Physics and Biruni Observatory, School of Science, Shiraz University, Shiraz 71454, Iran}}\\
\textit{{\small $^b$ School of Physics, Institute for Research in Fundamental Sciences (IPM), P. O. Box 19395-5531, Tehran, Iran}}\\
\textit{{\small $^c$ School of Particles and Accelerators, Institute for Research in Fundamental Science (IPM),}}\\
\textit{{\small P.O. Box 19395-5746, Tehran, Iran}}\\
\textit{{\small $^d$ Steklov Mathematical Institute, Russian Academy of Sciences, Gubkina str. 8, 119991, Moscow, Russia}}\\
\textit{{\small $^e$ Canadian Quantum Research Center 204-3002 32 Ave Vernon, BC V1T 2L7, Canada}}\\
}
\maketitle
\begin{abstract}
In this paper, we obtain analytical approximate black hole solutions in the framework of $f(R)$ gravity and the absence of a cosmological constant. In this area, we apply the equations of motion of the theory to a spherically symmetric spacetime with one unknown function and derive black hole solutions without any constraints on the Ricci scalar. To do so, first, we obtain the near horizon and asymptotic solutions and then use both of them to obtain a complete solution by utilizing a continued-fraction expansion. Finally, we investigate the stability of the solutions by employing the thermodynamics and quasi-normal modes. 
\end{abstract}

\section{Introduction}
~~~~More than a century ago, Newtonian gravity failed to explain the anomalies in the orbits of Mercury. In 1915, Einstein introduced his theory of General Relativity (GR), which not only described the answer to the physics problems of that time, including the issue of Mercury but also its ability in its interesting predictions like the gravitational lensing, gravitational waves, black holes, etc.
However, despite all of these successes, in recent years this theory, failed to be consistent with observation and was not able to describe the dark energy and dark matter that are confirmed by observations. Also, GR was not renormalizable and, therefore, could not be quantized by using the conventional methods in quantum field theory. It was shown that if one demands the renormalization to be satisfied at one-loop, the Einstein-Hilbert action should be supplemented by higher-order curvature terms \cite{Stelle:1976gc}.  Later on, Einstein's theory of GR had many forms of modifications such as: $f(R)$ gravity with $R$ as the Ricci scalar \cite{Nojiri:2010wj}, \cite{Capozziello:2011et}, \cite{Clifton:2011jh}, $f(R, T)$ gravity, where $T$ is the trace of energy-momentum tensor \cite{Harko:2011kv}, \cite{Zubair:2015gsb}, $f(G)$ gravity with $G$ the Gauss-Bonnet invariant \cite{Cognola:2006eg} etc. All of these modified theories have received much attention to investigating the accelerated expansion of our universe, flat rotation curves of galaxies, wormhole behavior \cite{DeFelice:2010aj}, \cite{Nojiri:2006ri}, \cite{Bamba:2012cp}, \cite{Sajadi:2011oei}. One of the familiar extensions of GR is to include terms such as
\begin{equation}
f(R)=...+\dfrac{\beta_{2}}{R^{2}}+\dfrac{\beta_{1}}{R}+R-\Lambda +\alpha_{2} R^2+\alpha_{3} R^3+...
\end{equation}
where $\alpha_{i}$ and $\beta_{i}$ are constants, and $\alpha_{i}=\beta_{i}=0$ corresponds to the Einstein-Hilbert action. The special case $\alpha_{2}\neq 0, \alpha_{i} =0(i\neq 2)$, and $\beta_{i} = 0$ is $R^{2}$ gravity and known as the Starobinsky model \cite{Starobinsky:1980te}.
In $f(R)$ gravity, the interaction of spacetime and matter is different from GR.
In GR, gravity appears as the curvature of spacetime, where the source of this curvature is all forms of matter. In the absence of any mass or energy, spacetime can become completely flat.
What $f(R)$ gravity does is allow spacetime to act as a source of its curvature so that there can still be some curvature even if spacetime is empty of matter. Therefore, as the universe expands and the matter density decreases, some curvature may remain that is capable of driving cosmic acceleration.

Many applications have been developed in the framework of the $f(R)$ gravity such as early-time inflation \cite{Starobinsky:1980te}, \cite{Bamba:2008ja}, \cite{Motohashi:2017vdc}, cosmological phases \cite{Nojiri:2006gh}, \cite{Nojiri:2006be}, \cite{Amendola:2006we}, \cite{Hu:2007nk}, \cite{Appleby:2007vb}, \cite{Starobinsky:2007hu}, \cite{Appleby:2009uf}, gravitational wave detection \cite{Corda:2008si}, \cite{Corda:2010zza}, the stability of the solutions \cite{Bhattacharyya:2017tyc}, \cite{Ovgun:2018gwt}, \cite{Aragon:2020xtm} and other different issues \cite{Akbar:2006mq}. Interestingly, the duality between gravity and quantum field theory that has been introduced in \cite{Maldacena:1997re} and its applications considered in different theories and $f(R)$ gravity such as \cite{Witten:1998qj}-\cite{Bousder:2021qls}. In addition, the different solutions of $f(R)$ gravity have been studied. Among these solutions, the authors in \cite{Multamaki:2006zb} and \cite{Multamaki:2006ym} deduced the static spherically symmetric solutions by involving a perfect fluid.  The black hole solution with/without electric charge is presented in \cite{delaCruz-Dombriz:2012bni}. Many analytic spherically symmetric solutions are derived in \cite{Hendi:2011hxq}, \cite{Hendi:2014mba}, \cite{Hendi:2011eg}. 

In this work, following the papers of quadratic \cite{Sajadi:2020axg} and cubic gravity \cite{Sajadi:2022ybs}, we concentrated on the analytic approximate black hole solutions of $f(R)$ gravity using a continued-fraction expansion and their stability through Quasi-Normal Modes (QNMs) and thermodynamic.

The paper is organized as follows: In Section \ref{sectt2}, first, we give a brief review of $f(R)$ gravity and apply the field equations of the theory to static-spherically symmetric spacetime.
Then, to obtain the full black hole solution, we combine the asymptotic and near horizon solutions using a continued-fraction expansion. In Section \ref{sectt3}, using the QNMs we study the dynamical stability of the black hole solutions. Finally, in Section \ref{conc} we
give our concluding remarks.

\section{Basic equations} \label{sectt2}

The action of $f(R)$ gravity can be written as 
\begin{equation}\label{eq1}
I=\dfrac{1}{8\pi G}\int d^{4}x \sqrt{-g} f(R)~.
\end{equation}
The variation to action gives the vacuum field equations as
\begin{equation}\label{eqmotion}
E_{\mu \nu}=R_{\mu \nu}f_{R}-\dfrac{1}{2}g_{\mu \nu}f(R)+[g_{\mu \nu}\square -\nabla_{\mu}\nabla_{\nu}]f_{R}=0~, 
\end{equation}
where $\square$ is the d'Alembertian operator and $f_{R} =df/dR$. The trace of the field equations takes the form
\begin{equation}\label{eqtrace}
E=E^{\mu}_{\mu}=3\square f_{R}+Rf_{R}-2f=0~.
\end{equation}
Inserting Eq. \eqref{eqtrace} in Eq. \eqref{eqmotion} we get
\begin{equation}\label{eqqfR}
E_{\mu \nu}=R_{\mu \nu}f_{R}-\dfrac{1}{4}g_{\mu \nu}Rf_{R}+\dfrac{1}{4}g_{\mu \nu}\square f_{R}-\nabla_{\mu}\nabla_{\nu}f_{R}=0~.
\end{equation}
In this paper, we consider $f(R)$ as follows
\begin{equation}\label{eqqfReq}
f(R)=R+\alpha R^{2}~.
\end{equation} 
Inserting into the field equation (\ref{eqqfR}), one can get
\begin{align}\label{eqqfieldequation}
R_{\mu \nu}-\dfrac{1}{4}g_{\mu \nu}R+\alpha\left[2RR_{\mu \nu}-\dfrac{1}{2}g_{\mu \nu}R^{2}+\dfrac{1}{2}g_{\mu \nu}\square R-2\nabla_{\mu}\nabla_{\nu}R\right]=0~.
\end{align}
For this model $\alpha>0$ from the Dolgov and Kawasaki stability point of view \cite{Dolgov:2003px}.
We consider the following spherically symmetric and static line element for describing
the geometry of spacetime
\begin{equation}\label{metform}
ds^{2}=-h(r)dt^{2}+\dfrac{dr^{2}}{h(r)}+r^2\left(d\theta^{2}+\sin^{2}(\theta)d\phi^{2}\right)~.
\end{equation}
As we know, generic static, spherically, and symmetric metrics do not need to obey $g_{tt}g_{rr} = -1$ necessarily, but field equation \eqref{eqqfieldequation} admits solutions with this property \cite{Nashed:2020mnp}, \cite{Nashed:2021lzq}, to which case we shall restrict our consideration here.
By inserting the metric into the field equations \eqref{eqqfieldequation}, the differential equations for $ h(r) $ become
\begin{align}\label{eq6}
E_{r r}=&-30\alpha r^2 h h^{\prime\prime}-24\alpha r h h^{\prime}+\alpha r^4h^{\prime}h^{\prime\prime\prime}+3\alpha h r^4 h^{\prime\prime\prime\prime}+10\alpha r^3 h h^{\prime\prime\prime}+4\alpha r^2h^{\prime\prime}-18\alpha r^{2}h^{\prime 2}\nonumber\\
&+28\alpha r h^{\prime}+2r^3 h^{\prime}+2r^2 h-28\alpha h+36\alpha h^2-8\alpha -2r^2=0~.
\end{align}
{We would like to emphasize that, utilizing the other components of field equations gives the same result.}
Expanding the function $h(r)$ around the event horizon $ r_{+} $ we obtain
\begin{align}\label{eq7}
h(r)  &= h_{1}(r-r_{+})+h_{2}(r-r_{+})^{2}+h_{3}(r-r_{+})^{3}+...  ~~,
\end{align}
and then inserting these expressions into equations (\ref{eq6}), we find
\begin{align}\label{eq9}
{h_{3}}=&-\dfrac{-4\alpha +h_{1}r_{+}^3+4\alpha r_{+}^2h_{2}-r_{+}^2+14\alpha r_{+}h_{1}-9\alpha r_{+}^2h_{1}^2}{3\alpha r_{+}^4h_{1}},\nonumber\\
 {h_{4}}=&\dfrac{24\alpha h_{1}h_{2}r_{+}^3-4h_{1}r_{+}^3+80\alpha r_{+}^2h_{1}^2-48\alpha r_{+}^3h_{1}^3+5r_{+}^4h_{1}^2-8\alpha +4\alpha r_{+}^2h_{2}-2r_{+}^2+24\alpha r_{+}^4h_{1}^2 h_{2}}{24\alpha r_{+}^6 h_{1}^2}\nonumber\\
 &+\dfrac{4\alpha h_{2}^2-h_{2}}{24\alpha r_{+}^2 h_{1}^2}~,
\end{align}
where $r_+$, $h_1$ and $h_{2}$ are undetermined constants of integration. The other near horizon constants are provided in the Appendix \ref{app2}.
In the large $r$ limit, we linearize the field equations about the flat background
\begin{equation}
h(r)\approx 1+\epsilon H(r)~,
\end{equation}
where $H(r)$ is to be determined by the field equations, and we linearize the differential equation by keeping terms only to order $\epsilon$. The resulting differential equation for $H(r)$ takes the form
\begin{equation}
H^{\prime\prime\prime\prime}+\dfrac{10}{3r}H^{\prime\prime\prime}-\dfrac{26}{3r^{2}}H^{\prime\prime}+\dfrac{2(r^2+2\alpha)}{3\alpha r^{3}}H^{\prime}+\dfrac{2(r^2+22\alpha)}{3\alpha r^{4}}H=0~.
\end{equation}
In the large $r$ limit we obtain
\begin{equation}
H^{\prime\prime\prime\prime}+\dfrac{10}{3r}H^{\prime\prime\prime}-\dfrac{26}{3r^{2}}H^{\prime\prime}+\dfrac{2}{3\alpha r}H^{\prime}+\dfrac{2}{3\alpha r^{2}}H=0~,
\end{equation}
that can be solved as
{ \begin{align}
H(r)=&c_{1}r^{4}\, _0F_2 \left(;3,\dfrac{25}{6};-\dfrac{r^{2}}{12\alpha}\right)+c_{2}r^{-\frac{7}{3}} \, _0F_2 \left(;-\dfrac{13}{6},-\dfrac{1}{6};-\dfrac{r^{2}}{12\alpha}\right)\nonumber\\
&+c_{3}r \, _1F_3 \left(1;-\dfrac{1}{2},\dfrac{2}{3},\dfrac{8}{3};-\dfrac{r^{2}}{12\alpha}\right)~,
\end{align}}
here, the $pFq(a,b,z)$ is the generalized hypergeometric function \cite{Gradshteyn}, \cite{maple}.
For $c_{1}=c_{2}=c_{3}=H_{1}$ and in the large $r$, $H(r)$ becomes
\begin{align}
H(r)=&\dfrac{H_{1}}{r}-\dfrac{10\alpha H_{1}}{r^{3}}-\dfrac{210\alpha^{2}H_{1}}{r^{5}}-\dfrac{37800\alpha^{3}H_{1}}{r^{7}}-\dfrac{20374200\alpha^{4}H_{1}}{r^{9}}-\dfrac{23837814000\alpha^{5}H_{1}}{r^{11}}\nonumber\\
&+\mathcal{O}\left(\dfrac{1}{r^{12}}\right)~.
\end{align}
Finally, the solution is
\begin{equation}\label{eq70} 
h(r)=1+H(r)~.
\end{equation}
We wish to obtain an approximate analytic solution that is valid near the horizon and at large $r$. To reach this goal, we employ a continued-fraction expansion and write \footnote{{It should be noted that the other components of field equations give the same result for the equation \eqref{eq7}-\eqref{eq70}.}} \cite{Kokkotas:2017zwt}-\cite{Rezzolla:2014mua}
\begin{equation}\label{eq17}
h(r)=xA(x),\hspace{1cm} x= 1- \frac{r_+}{r}~,
\end{equation}
with
\begin{align}
A(x) &=1-\epsilon(1-x)+(a_{0}-\epsilon)(1-x)^{2}+\dfrac{a_{1}(1-x)^{3}}{1+\dfrac{a_{2}x}{1+\dfrac{a_{3}x}{1+\dfrac{a_{4}x}{1+...}}}}~.
\label{Ax}
\end{align} 
{Even at lowest order, the continued fraction approximation does a good job of approximating the solution everywhere outside the horizon. This only gets better as more terms are included.} 
Here, we truncated the continued-fraction expansion at order $4$. {This method give an accurate analytic expression approximating for the metric for the whole space outside the event horizon, and not only near the black hole or far from it $(r_+\leq r<\infty)$.}
By expanding (\ref{eq17}) near the horizon ($ x\to 0 $) and 
the asymptotic  region ($ x\to 1 $)  we obtain  
\begin{equation}
\epsilon=-\dfrac{H_{1}}{r_{+}}-1, \qquad a_{0}=0,\qquad a_{1}=-1-a_{0}+2\epsilon+r_{+}h_{1}~,
\end{equation}
for the lowest order expansion coefficients, with the remaining
$a_i$ given in terms of $(r_+, h_1, h_2)$. Also, we provided these expressions in the  Appendix \ref{appa}.  
 
\begin{figure}[H]\hspace{0.4cm}
\centering
\subfigure[]{\includegraphics[width=0.3\columnwidth]{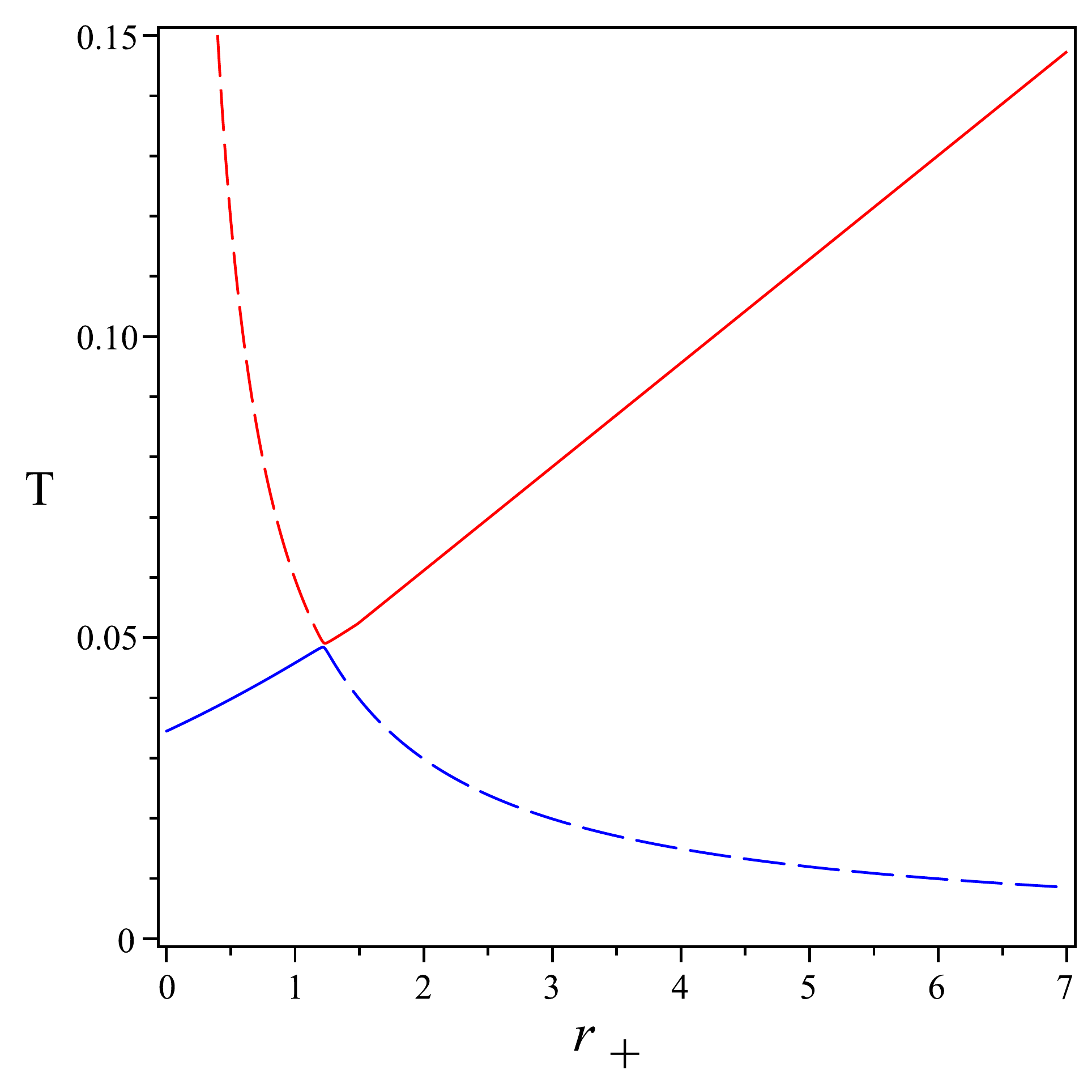}}
\subfigure[]{\includegraphics[width=0.3\columnwidth]{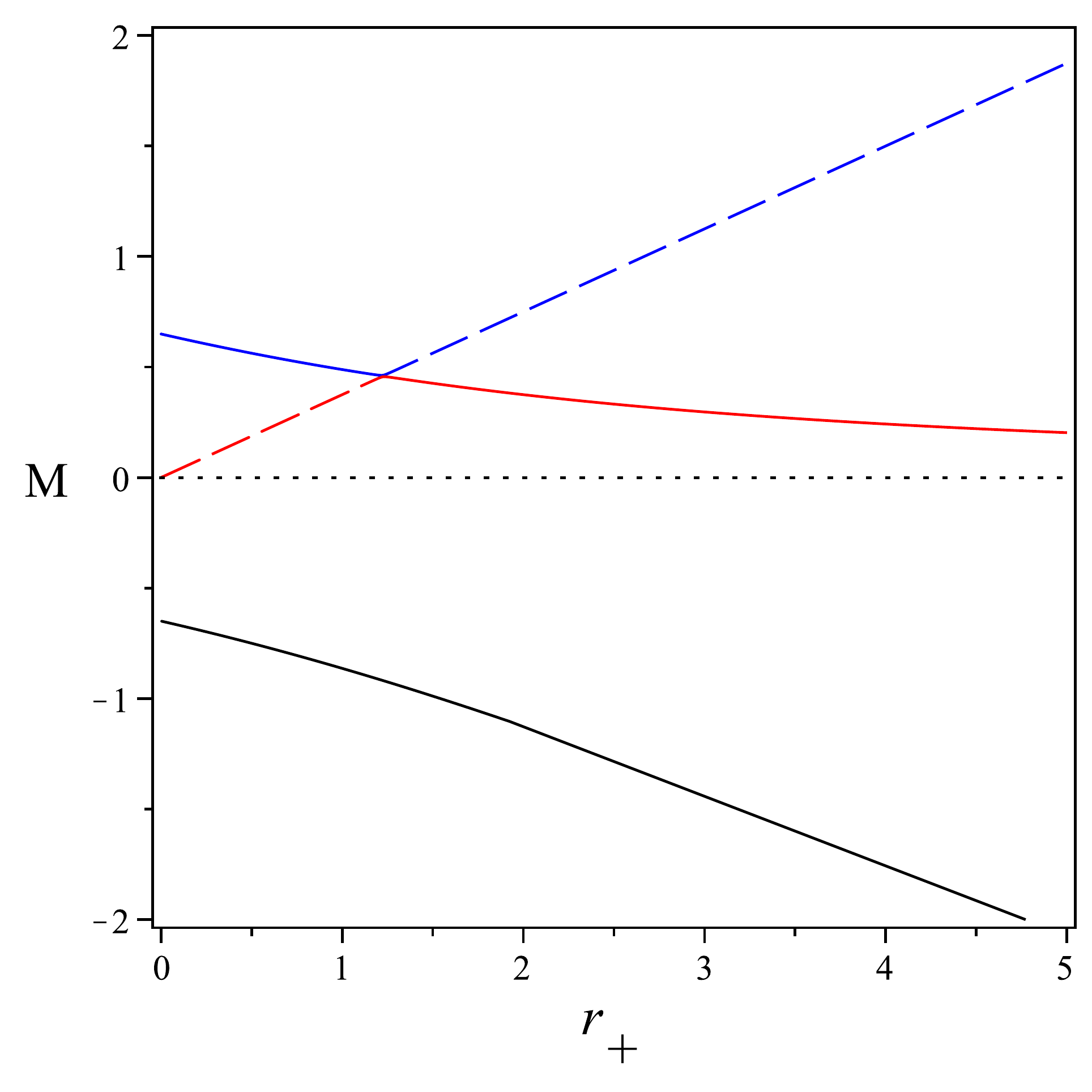}}
\subfigure[]{\includegraphics[width=0.3\columnwidth]{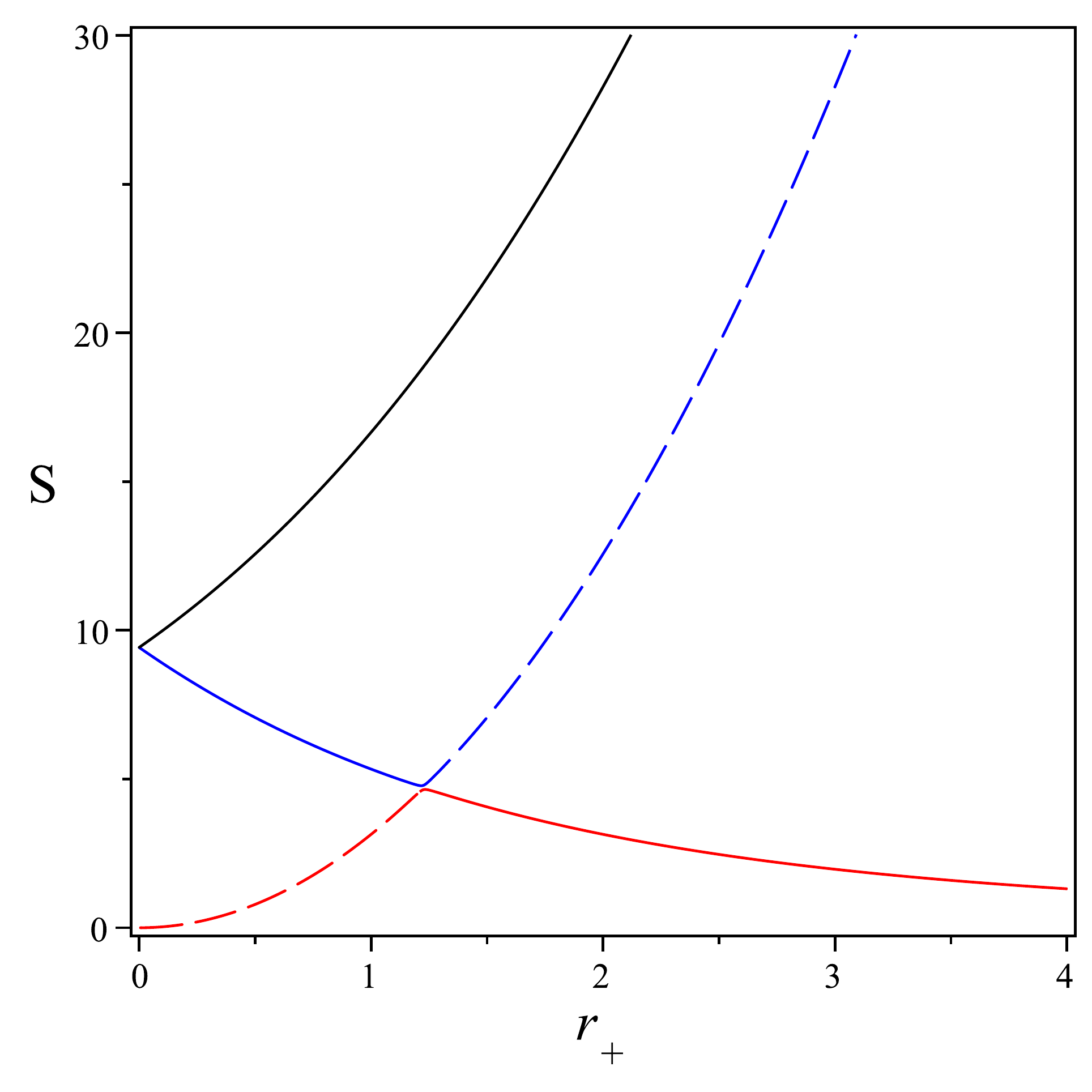}}
\caption{Plots of $T$, $M$ and $S$ in terms of $r_+$ for $\alpha=0.5,c_{1}=-0.1406$. In all panels the dashed line curves indicate Schwarzschild-like behavior and solid line curves non-Schwarzschild-like behavior. } 
\label{TMSplote}
\end{figure}

\begin{figure}[H]\hspace{0.4cm}
\centering
\subfigure[]{\includegraphics[width=0.3\columnwidth]{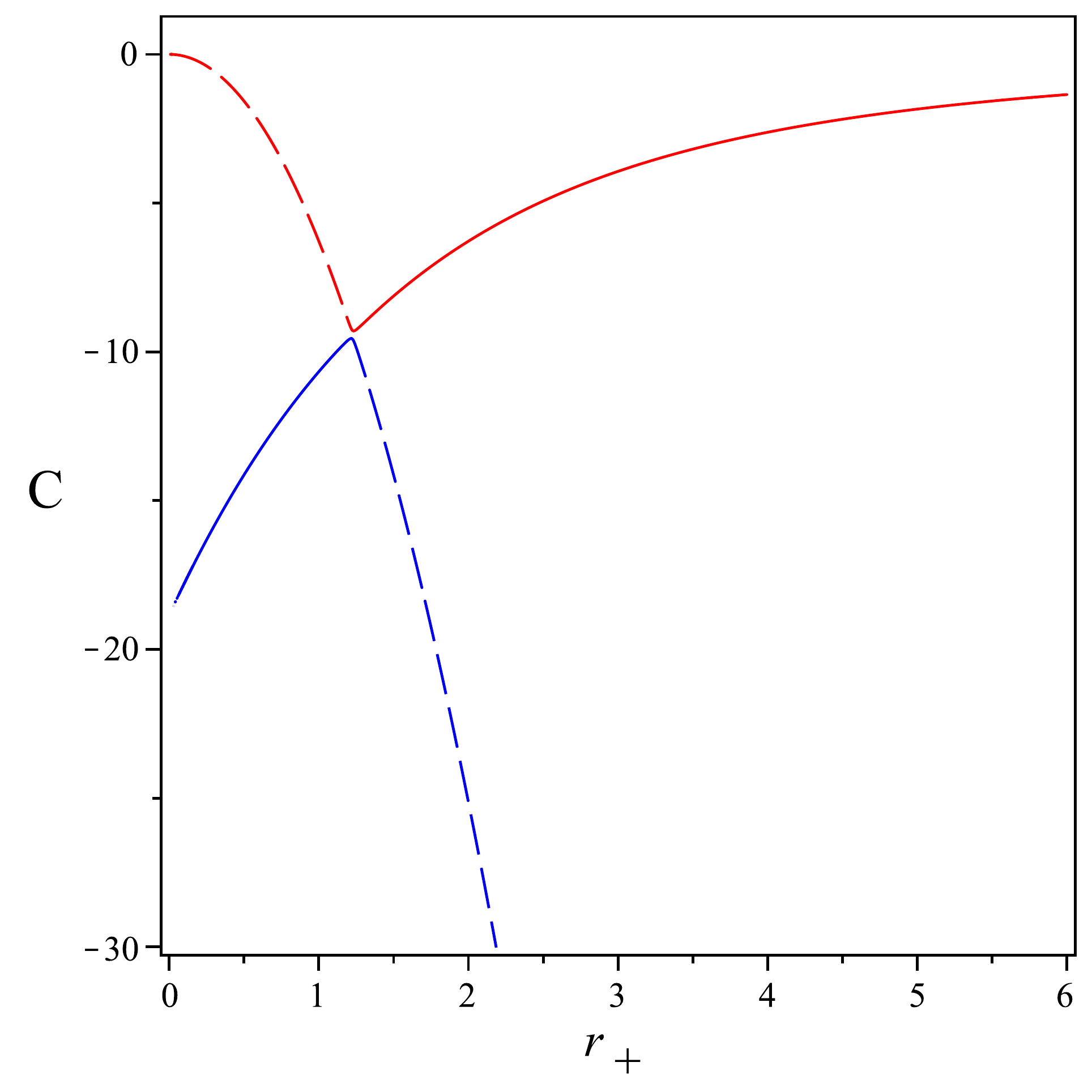}}
\subfigure[]{\includegraphics[width=0.3\columnwidth]{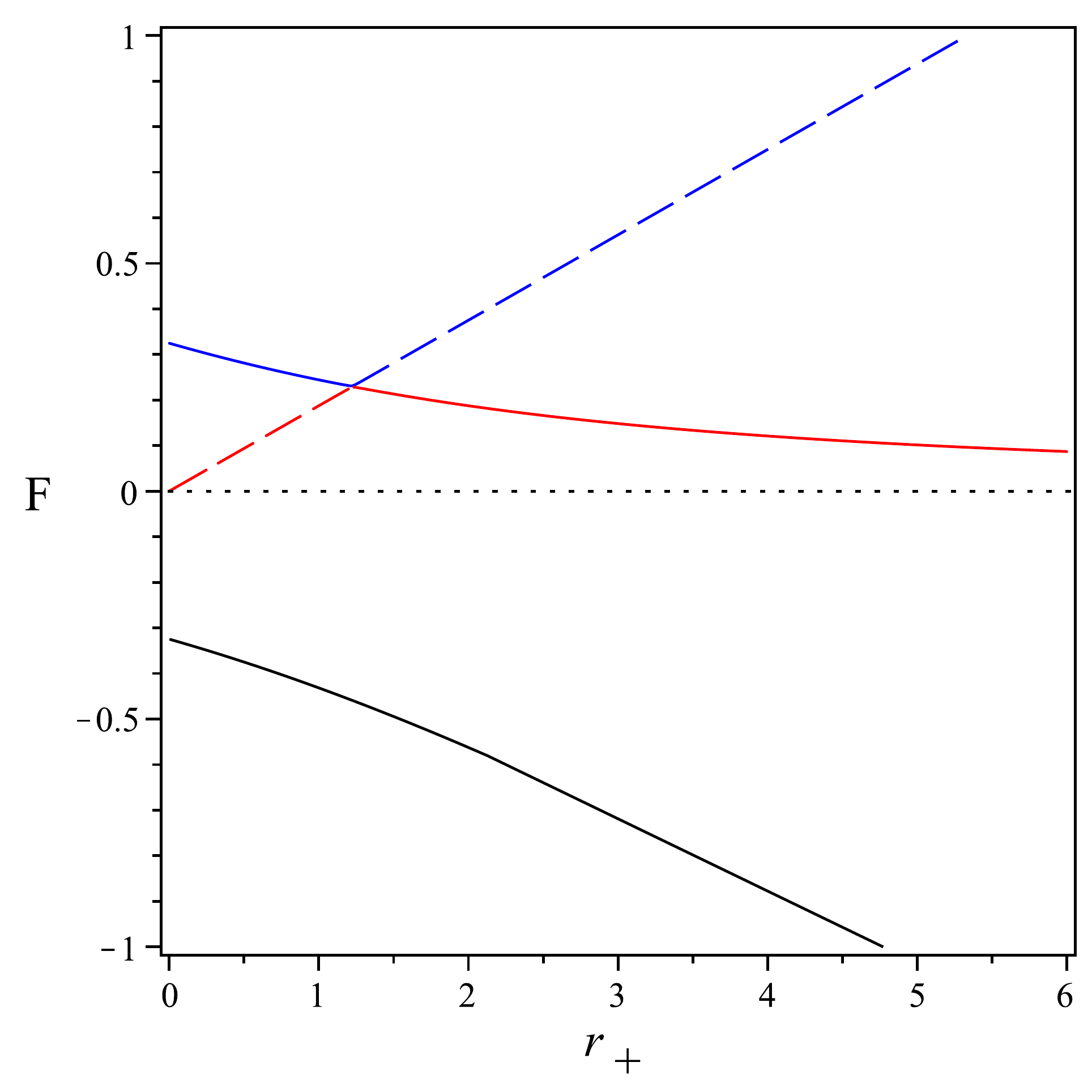}}
%\subfigure[$r_{+}=0.5$]{\includegraphics[width=0.3\columnwidth]{ffrrplot21}}
\caption{The behavior of heat capacity (left) and free energy (right) in terms of $r_+$ for $\alpha=0.5,c_{1}=-0.1406$.} 
\label{FCplote}
\end{figure}

The result is an approximate analytic solution for metric functions everywhere outside the horizon. For a static space time 
we have a timelike Killing vector 
$ \xi=\partial_{t} $ everywhere outside the horizon. Therefore, one can obtain
\begin{align}\label{eq20}
T &=\left. \dfrac{h^{'}(r)}{4\pi}\right\vert_{r_{+}}
= \dfrac{h_{1}}{4\pi} = {\dfrac {(1-2\epsilon+a_{1}+a_{0})}{{4\pi r_{+}}}}~.
 \end{align}
We computed the entropy as follows \cite{Wald1}, \cite{Wald2} 
\begin{align}\label{eqqqentropy}
S=&\dfrac{\mathcal{A}}{4}f_{R}=\dfrac{\mathcal{A}}{4}\left[1-\dfrac{2\alpha}{r^2} \left(h^{\prime\prime }r^2+4r h^{\prime}-2+2h\right)\right]_{r=r_{+}}=\pi r_{+}^{2}\left[1-\dfrac{2\alpha}{r_{+}^2} \left(h_{2}r_{+}^2+4r_{+}h_{1}-2\right)\right].
\end{align}
 We now consider the thermodynamics of these black hole solutions, whose basic equations are 
the first law and Smarr formula 
\begin{equation}\label{eqfirstlaw}
dM=TdS~,
\end{equation}
\begin{equation}\label{eq26}
M=2TS~,
\end{equation}
where there are no pressure/volume terms since we have set $\Lambda=0$. 
From Eq. \eqref{eq26} we have
\begin{equation}\label{eqmass}
M=\dfrac{h_{1}(r_{+})}{2}\left(r_{+}^2-2\alpha r_{+}^2h_{2}(r_{+})-8\alpha r_{+}h_{1}(r_{+})+4\alpha\right)~,
\end{equation}
yielding the mass parameter as a function of the horizon radius. 
Regarding the asymptotic flat spacetimes, we can define the total mass at spatial infinity as the Arnowitt-Deser-Misner (ADM) mass. Besides, the Komar mass makes sense for 
calculating total mass of a black hole solutions enjoying a timelike Killing vector. In addition to these two approaches, one can use alternative methods like Misner-Sharp 
prescription. Although the Misner-Sharp mass is just the Schwarzschild-like mass in the Schwarzschild space, it can be extended to 4-dimensional f(R) gravity (see for example: 
\cite{Cai:2009qf}, \cite{Cai:2008mh}, \cite{Zhang:2014goa} for higher dimensions).
In the Appendix \ref{massMS}, we showed that the asymptotic behavior of the mass (\ref{eqmass}) is the same as the mass of the Schwarzschild black hole. Therefore, one can interpret the mass (\ref{eqmass}) as ADM mass. 
 We now impose the first law \eqref{eqfirstlaw}, which becomes
\begin{align}\label{flaw1}
\dfrac{\partial M}{\partial r_{+}}-T\dfrac{\partial S}{\partial r_{+}}=0~,
\end{align}
where differential equation that must be satisfied by $h_{1}(r_{+})$ and $h_{2}(r_{+})$. Therefore, we achieve the prominent equation as follows 
\begin{align}\label{eqqfield1}
[&r_{+}^2-2r_{+}^2\alpha h_{2}(r_{+})-12\alpha r_{+}h_{1}(r_{+})+4\alpha]h^{\prime}_{1}(r_{+})-h_{1}(r_{+})[4\alpha h_{1}(r_{+})-r_{+}\nonumber\\
&+2\alpha r_{+}h_{2}(r_{+})+\alpha r_{+}^2h^{\prime}_{2}(r_{+})]=0~.
\end{align}
In order to solve the differential equation \eqref{eqqfield1}, we need to define $h_{2}$. To do so, we assume $h_{2}=g(\alpha)$. Recall that, near the horizon, the metric function is expanded as
\begin{equation}
h(r)=h_{1}(r-r_{+})+h_{2}(r-r_{+})^{2}+\sum h_{i}(h_{2})(r-r_{+})^{i}~,
\end{equation}
where the constants with $i >2$ are determined by the field equations in terms of other parameters. We will demand that this expansion has a smooth $\alpha\to 0$ limit. Therefore, the expansions for the first four terms are
\begin{align}
h_{3}(h_{2})=&-\dfrac{r_{+}h_{1}-1}{3h_{1}r_{+}^{2}\alpha}-\dfrac{14h_{1}r_{+}-9h_{1}^{2}r_{+}^{2}-4+4r_{+}^{2}g(0)}{3h_{1}r_{+}^{4}}+\mathcal{O}(\alpha),\\
h_{4}(h_{2})=&\dfrac{5r_{+}^{4}h_{1}^{2}-2r_{+}^{2}-4r_{+}^{3}h_{1}-r_{+}^{4}g(0)}{24h_{1}^{2}r_{+}^{6}\alpha}+
\dfrac{4r_{+}^{4}g(0)^{2}-48h_{1}^{3}r_{+}^{3}
+4r_{+}^{2}g(0)}{24h_{1}^{2}r_{+}^{6}}\nonumber\\
&+\dfrac{24r_{+}^{4}h_{1}^{2}g(0)+24r_{+}^{3}h_{1}g(0)-r_{+}^{4}g^{\prime}(0)+80h_{1}^{2}r_{+}^{2}-8}{24h_{1}^{2}r_{+}^{6}}+\mathcal{O}(\alpha)~,\\
h_{5}(h_{2})=&\dfrac{h_{1}r_{+}-1}{210h_{1}^{3}r_{+}^{4}\alpha^{2}}+
\dfrac{100h_{1}^{2}r_{+}^{2}-12-
3h_{1}^{2}r_{+}^{4}g(0)+5r_{+}^{4}g(0)^{2}+2r_{+}^{3}h_{1}g(0)}{420h_{1}^{3}r_{+}^{6}\alpha}\nonumber\\
&+\dfrac{28r_{+}^{2}g(0)+80r_{0}h_{1}-146h_{1}^{3}r_{+}^{3}}{420h_{1}^{3}r_{+}^{6}\alpha}+\mathcal{O}(\alpha)~,\\
h_{6}(h_{2})=&-\dfrac{90r_{+}^{7}g(0)h_{1}-24h_{1}r_{+}^{5}-97r_{+}^{6}g(0)-106r_{+}^{4}
+165r_{+}^{6}h_{1}^{2}-42r_{+}^{7}h_{1}^{3}}{25200h_{1}^{4}r_{+}^{10}\alpha^{2}}\nonumber\\
&-\dfrac{120g(0)^{3}r_{+}^{8}-97r_{+}^{6}g^{\prime}(0)-72g(0)^{2}r_{+}^{8}h_{1}^{2}+90r_{+}^{7}g^{\prime}(0)h_{1}+1976r_{+}^{3}h_{1}}{25200r_{+}^{10}h_{1}^{4}\alpha}\nonumber\\
&+\dfrac{4200r_{+}^{4}h_{1}^{2}+6420h_{1}^{3}r_{+}^{5}+60r_{+}^{7}h_{1}^{3}g(0)+3692r_{+}^{5}h_{1}g(0)+216g(0)r_{+}^{4}}{25200r_{+}^{10}h_{1}^{4}\alpha}\nonumber\\
&+\dfrac{1084r_{+}^{6}g(0)^{2}-
832r_{+}^{2}-9882r_{+}^{6}h_{1}^{4}-1374r_{+}^{6}g(0)h_{1}^{2}-18
0h_{1}r_{+}^{7}g(0)^{2}}{25200r_{+}^{10}h_{1}^{4}\alpha}+\mathcal{O}(\alpha^{0})~.
\end{align}
Demanding  $h_{3}$ to have a smooth behavior at $\alpha\to 0$ limit, we must take
\begin{equation}\label{eqqpadap1}
h_{1}=\dfrac{1}{r_{+}}~.
\end{equation}
Then, for $h_{4}$ we have
\begin{equation}\label{eqqpadap2}
g(0)=-\dfrac{1}{r_{+}^{2}}~.
\end{equation}
Now, by inserting (\ref{eqqpadap1}) and (\ref{eqqpadap2}) into the $h_{5}$ and $h_{6}$, one can find $h_{5}=0$ and $g^{\prime}=0$. 
By continuing the procedure in this way, one can find that the other coefficients of expansion should be zero.
\begin{equation}
g^{\prime\prime}(0)=g^{\prime\prime\prime}(0)=g^{\prime\prime\prime\prime}(0)...=0~.
\end{equation}
Therefore, $h_{2}$ can be obtained as
\begin{equation}\label{eqq36q}
h_{2}=g(0)+g^{\prime}\alpha+g^{\prime\prime}\alpha^{2}+...=-\dfrac{1}{r_{+}^{2}}~.
\end{equation}
By inserting $h_{2}$ into the equation (\ref{eqqfield1}), one can obtain a differential equation only for $h_{1}$. By solving it we obtained three {solutions}  for $h_{1}$ 
\begin{align}\label{eqqh1}
h_{1}^{(1)}&=\dfrac{B^{\frac{1}{3}}}{24\alpha r_{+}}+\dfrac{(r_{+}^{2}+6\alpha)^{2}}{24\alpha r_{+}B^{\frac{1}{3}}}+\dfrac{r_{+}^{2}+6\alpha}{24\alpha r_{+}}~,\\
h_{1}^{(2)}&=-\dfrac{B^{\frac{1}{3}}}{48\alpha r_{+}}-\dfrac{(r_{+}^{2}+6\alpha)^{2}}{48\alpha r_{+}B^{\frac{1}{3}}}+\dfrac{r_{+}^{2}+6\alpha}{24\alpha r_{+}}-\dfrac{i\sqrt{3}}{2}\left(\dfrac{B^{\frac{1}{3}}}{24\alpha r_{+}}-\dfrac{(r_{+}^{2}+6\alpha)^{2}}{24\alpha r_{+}B^{\frac{1}{3}}}\right)~,\\
h_{1}^{(3)}&=-\dfrac{B^{\frac{1}{3}}}{48\alpha r_{+}}-\dfrac{(r_{+}^{2}+6\alpha)^{2}}{48\alpha r_{+}B^{\frac{1}{3}}}+\dfrac{r_{+}^{2}+6\alpha}{24\alpha r_{+}}+\dfrac{i\sqrt{3}}{2}\left(\dfrac{B^{\frac{1}{3}}}{24\alpha r_{+}}-\dfrac{(r_{+}^{2}+6\alpha)^{2}}{24\alpha r_{+}B^{\frac{1}{3}}}\right)~,
\end{align}
where
\begin{align}
A&=c_{1}(216\alpha^3 +1728c_{1}\alpha^2 r_{+}^{2}+r_{+}^{6}+18\alpha r_{+}^{4}+108\alpha^{2}r_{+}^{2})\\
B&=3456c_{1}\alpha^2 r_{+}^{2}+r_{+}^{6}+18\alpha r_{+}^{4}+108\alpha^{2}r_{+}^{2}+216\alpha^{3}+48\alpha r_{+}\sqrt{3A}.\label{eqqab}
\end{align}
{For $\alpha>0$ and $c_1>0$ we have just one real solution and also for $\alpha>0$ and $c_1 \leq 0$ we have three real solutions.} Inserting (\ref{eqqh1})-(\ref{eqqab}) into the thermodynamical quantities and plotting them, one can obtain Fig. \eqref{TMSplote}. In this figure, we have illustrated the temperature, mass, and entropy for different values of parameters in the left, middle, and right panels, respectively. As can be seen, there are two kinds of solutions with four branches in such a way that the black dashed lines are the Schwarzschild-like solution and the solid lines non-Schwarzschild-like solution. The black solid line is a non-physical black hole solution due to having a negative temperatures and mass.
In figure \eqref{FCplote} we have shown the behavior of heat capacity and free energy in the left and right panels, respectively. As can be seen, the heat capacity is negative for all values of $r_{+}$ which shows the black holes locally are unstable. The free energy for a non-Schwarzschild-like solution is decreasing function which shows the solutions globally are stable.

\begin{figure}[H]\hspace{0.4cm}
\centering
\subfigure[$r_{+}=2, m=0.37$]{\includegraphics[width=0.3\columnwidth]{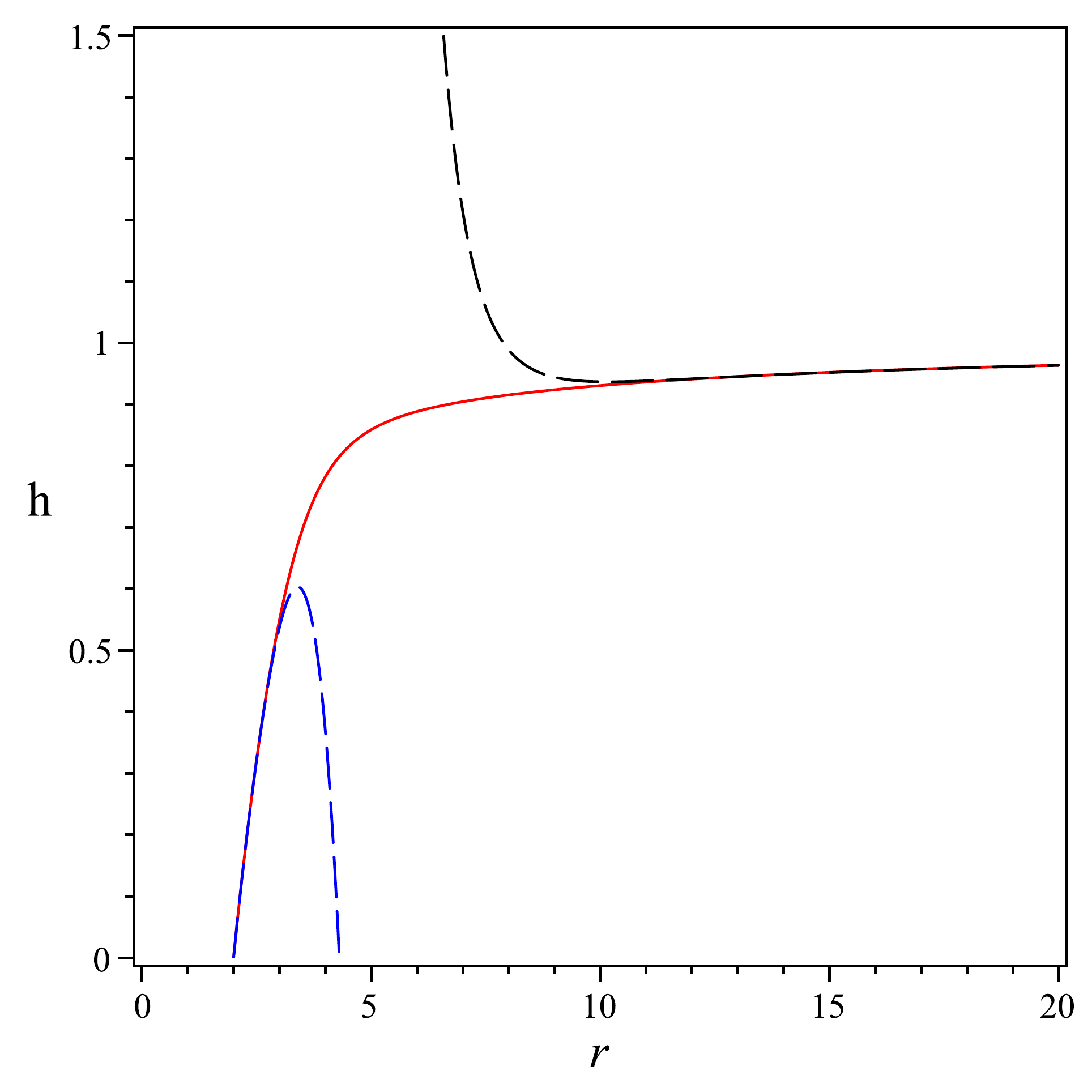}}
\subfigure[$r_{+}=1.5, m=0.57$]{\includegraphics[width=0.3\columnwidth]{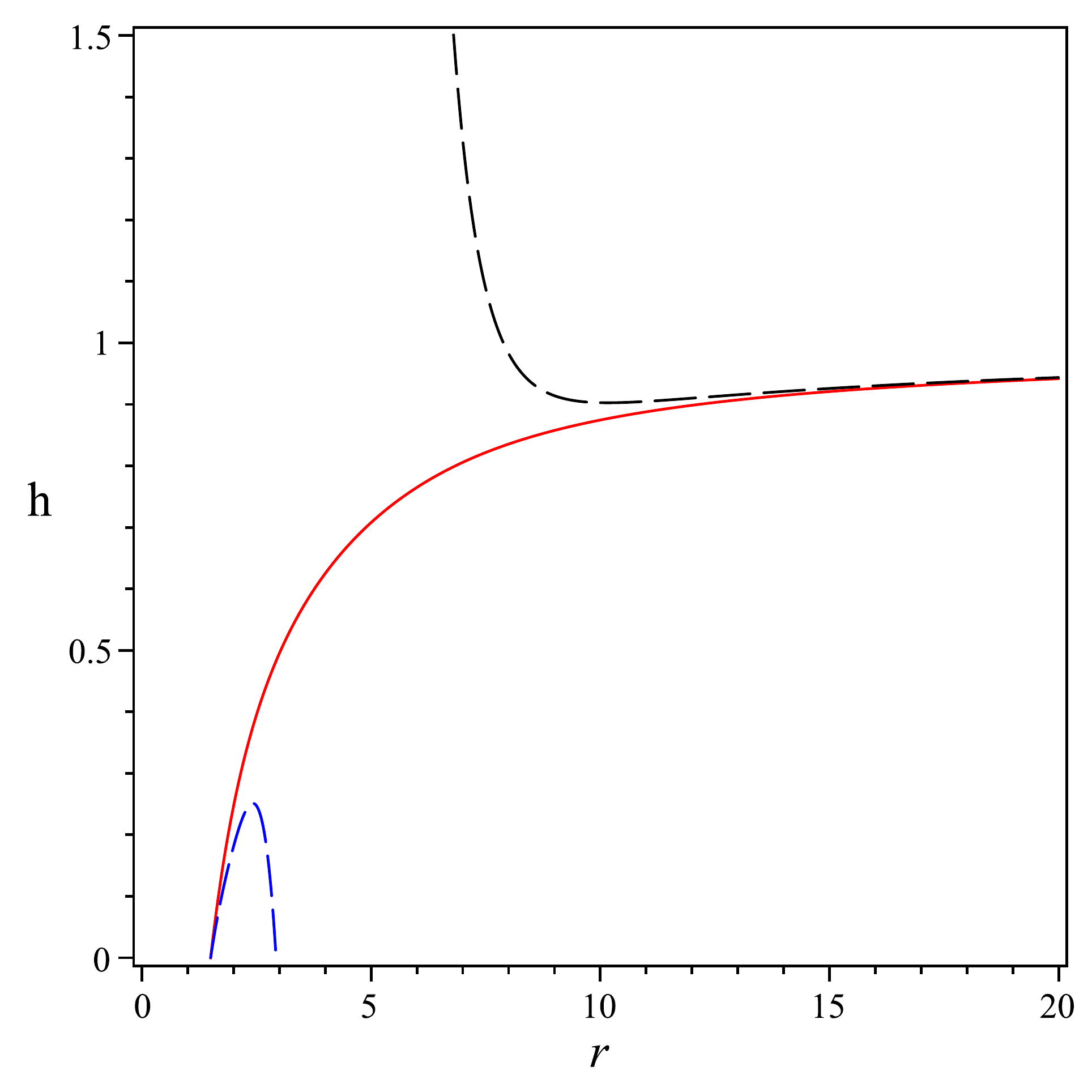}}
\subfigure[$r_{+}=0.5, m=0.55$]{\includegraphics[width=0.3\columnwidth]{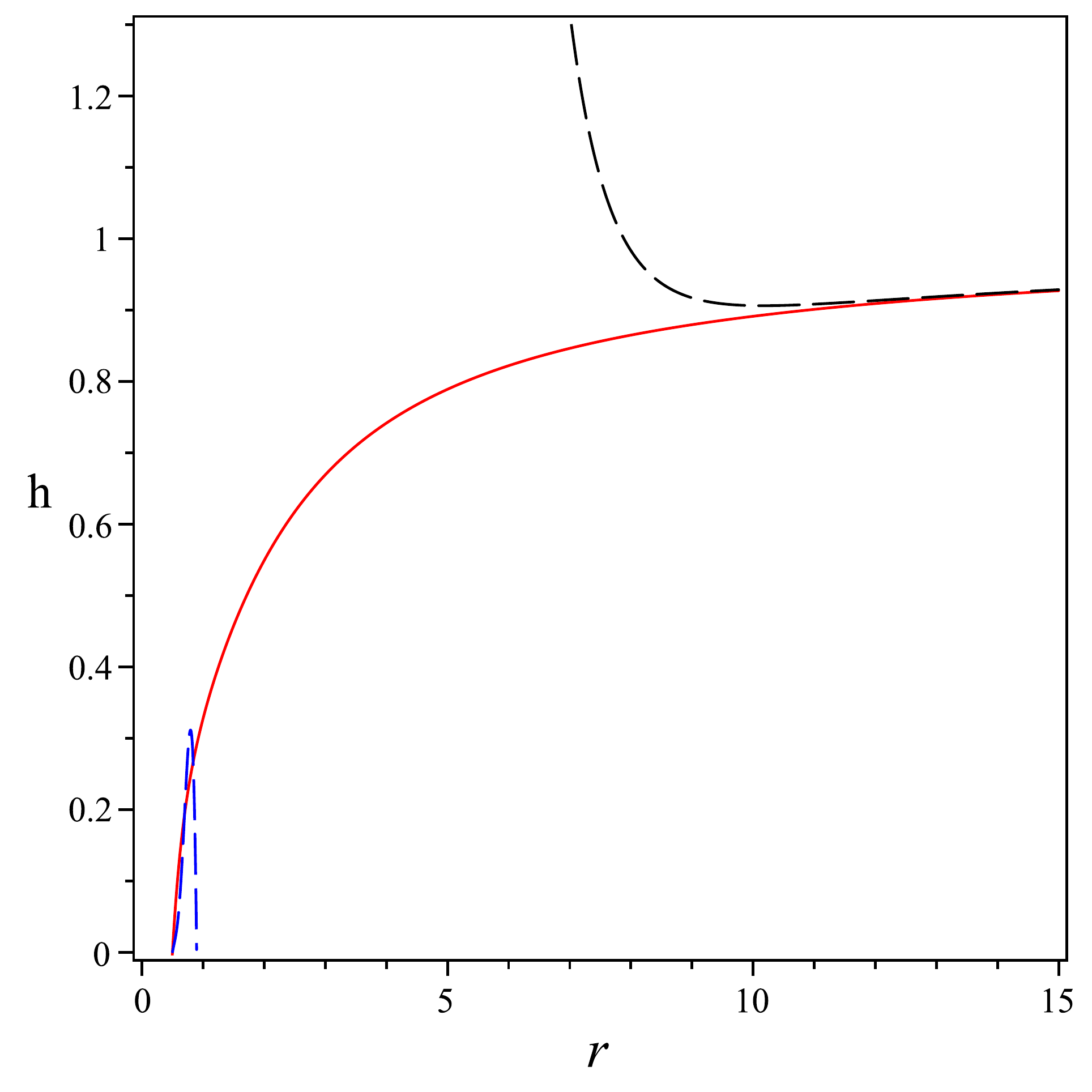}}
\caption{The plots of metric for $\alpha=0.5,c_{1}=-0.1406$. The red solid line is the full continued-fraction solution, the blue dashed line is the near-horizon solution and black dashed line is the large-$r$ solution.} 
\label{frplote}
\end{figure}

In Figure (\ref{frplote}) we present the solutions for $ h(r) $ with different values of horizons in the left, middle, and right panels that depicting the full continued-fraction solution (\ref{eq17}) along with its comparison to the near-horizon solution with blue dashed lines and large-$r$ 
 series expansions solution with black dashed lines.  We see that
the continued-fraction expansion converges to both of these two approximations. Similar to quadratic gravity, we find two groups of solutions \cite{Lu:2015cqa}, \cite{Lu:2015psa}, \cite{Sajadi:2020axg}. For the first group, the metric function is increasing functions in $ r\geq r_{+} $. These solutions are generalizations of the Schwarzschild black hole solution and reduce to the Schwarzschild as $\alpha\to 0$.  
The non-Schwarzschild solutions are physically distinct from the first group, and as $\alpha\to 0$, it does not go back to the Schwarzschild metric.

\begin{figure}[H]\hspace{0.4cm}
\centering
\subfigure[]{\includegraphics[width=0.45\columnwidth]{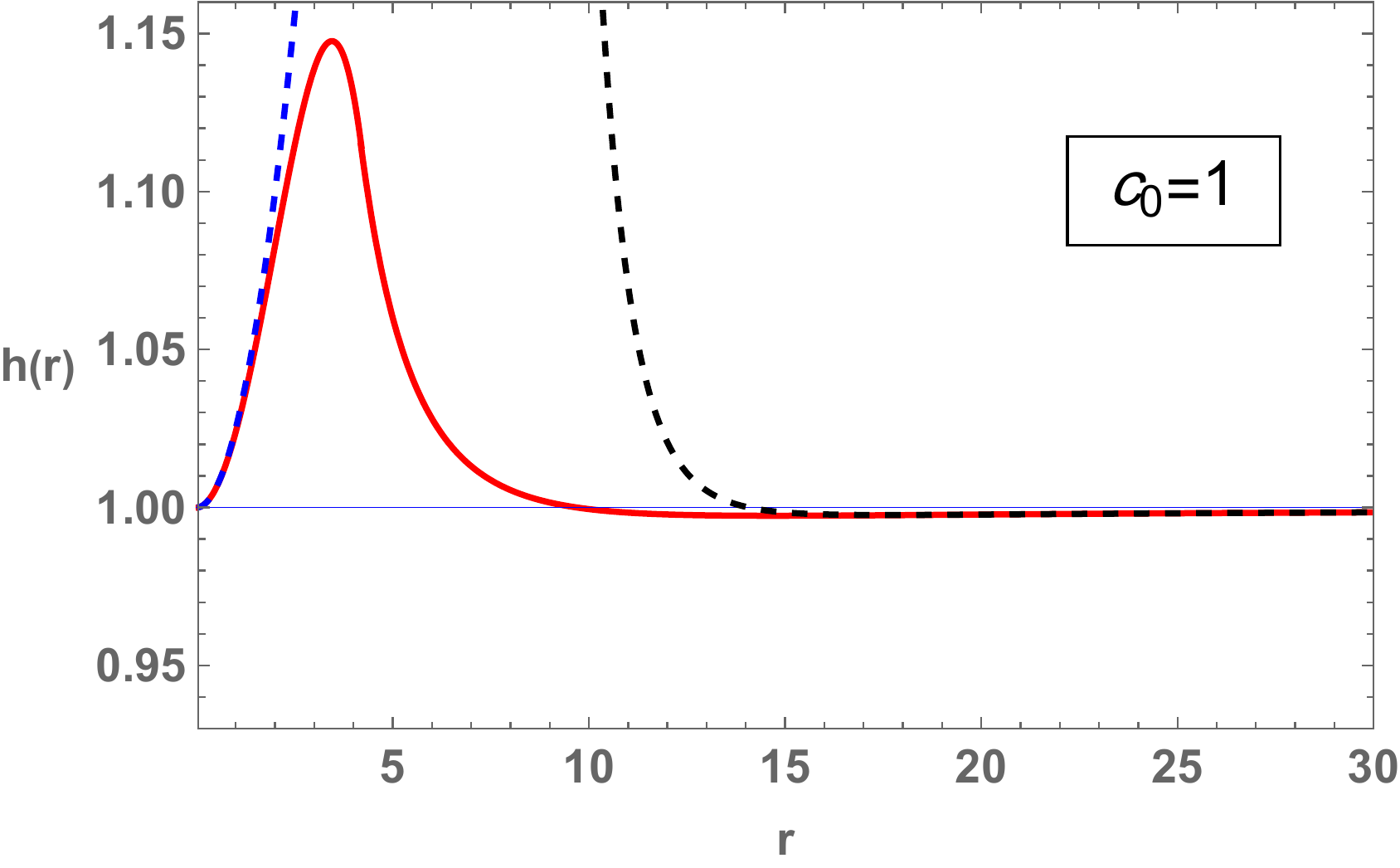}}
%\subfigure[]{\includegraphics[width=0.45\columnwidth]{}}
\subfigure[]{\includegraphics[width=0.45\columnwidth]{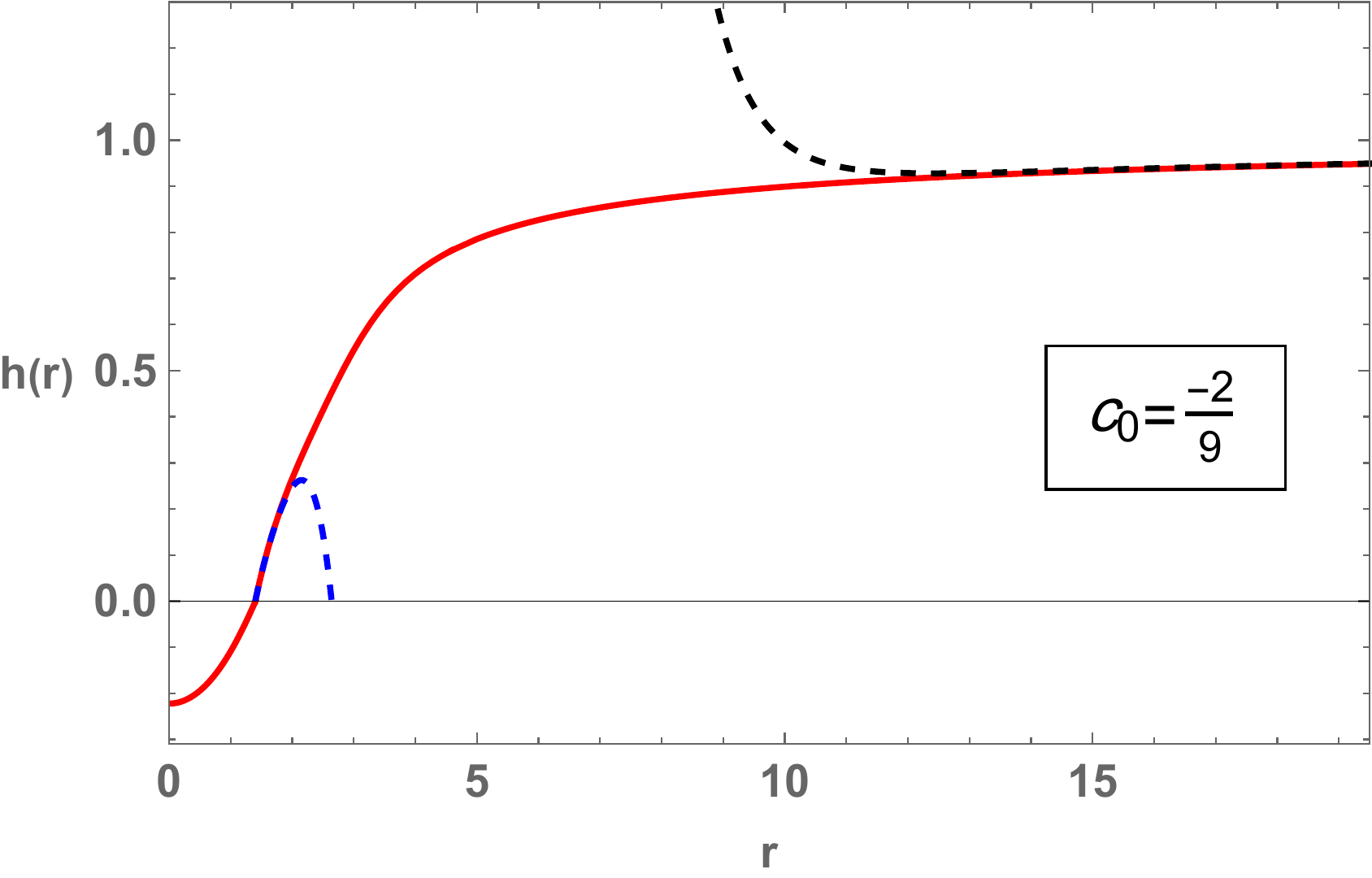}}
\caption{The solution without horizon (left panel) and Schwarzschild-like black hole solution with horizon radius $r_{+}=1.4$ (right panel) in terms of $r$ for $\alpha=1,H_1=-0.047, c_0=1$, and $\alpha=0.5,H_1=-1, c_0=\frac{-2}{9}$, respectively. At both cases we set $c_{1}=-0.1406$.} 
\label{nakedhr}
\end{figure}

Here, we  would like to solve the field equation (8) and the metric near the center of a black hole. Therefore, we consider the following expansion around the origin as
\begin{equation}
h(r)=\sum_{i=0} c_{i}r^{i}~.
\end{equation}
Similar to previous cases for near horizons and asymptotes, one can obtain the metric as
\begin{align}\label{nearoriginmetric}
h(r)=&c_{0}+\dfrac{c_{0}r^{2}}{4\alpha (9c_{0}+1)}+\dfrac{3(3c_{0}+1)c_{0}^{2}r^{4}}{8\alpha^{2}(9c_{0}+1)^{2}(27c_{0}^{2}-40c_{0}-2)}+\nonumber\\
&\dfrac{15c_{0}^{3}(9c_{0}^{2}-3c_{0}-2)r^{6}}{64\alpha^{3}(159c_{0}^{3}+36c_{0}+1)(9c_{0}+1)^{3}(27c_{0}^{2}-40c_{0}-2)}+\mathcal{O}(r^{8})~,
\end{align}
with $c_{0}=1$ and $c_{0}=-2/9$. The metric is fully determined and there is no free parameter.
The Kretschmann scalar {and Ricci scalars} near the origin behaves as
\begin{equation}
K=R_{a b c d}R^{a b c d}=\dfrac{4(c_{0}-1)^{2}}{r^{4}}+\dfrac{2c_{0}(c_{0}-1)}{\alpha (9c_{0}+1)r^{2}}+\dfrac{3c_{0}^{2}(33c_{0}^{2}-44c_{0}-4)}{2\alpha^{2}(9c_{0}+1)^{2}(27c_{0}^{2}-40c_{0}-2)}+\mathcal{O}(r^{2})~,
\end{equation}
\begin{equation}
{ R=\dfrac{2(1-c_{0})}{r^{2}}-\dfrac{2}{3\alpha}+\mathcal{O}(r^{2})~.}
\end{equation}
As can be seen, for the case $c_{0} = 1$, the behavior of the metric near the origin corresponds to an Anti-de Sitter space-time and the Kretschmann and Ricci scalar have a finite values ($K=3/200\alpha^2, R=-2/3\alpha$) (with a nonzero mass). Regular solutions with an Anti-de Sitter core have been studied with a nonlinear electromagnetic source in \cite{Balart:2014cga}. But, for the case $c_{0} = -2/9$, the Kretschmann and Ricci scalar do not become zero and there is a singularity.
In Figure \ref{nakedhr}, we have shown the full solutions of the field equations from center to infinity by starting from the near origin metric (\ref{nearoriginmetric}) and asymptotic solution \eqref{eq70}, for $c_0=1$ (a solution without horizon\footnote{This solution is not a naked solution because the Kretschmann and Ricci scalar do not diverge.}) and for $c_0=-2/9$ (black hole case).  

\section{Dynamical Stability}\label{sectt3}
\begin{figure}[H]\hspace{0.4cm}
\centering
\subfigure[$r_{+}=0.5,M=0.55$]{\includegraphics[width=0.3\columnwidth]{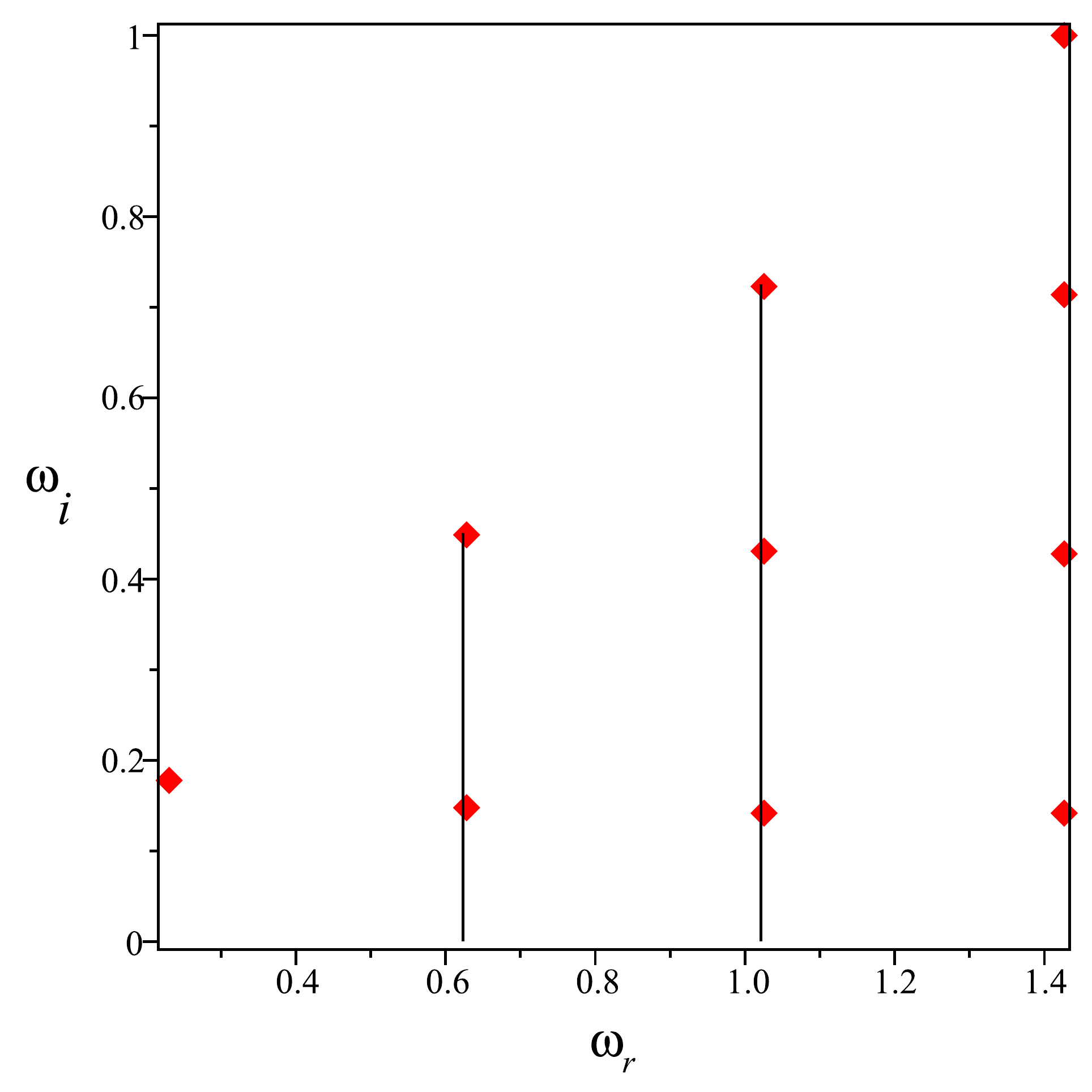}}
\subfigure[$r_{+}=1.5,M=0.57$]{\includegraphics[width=0.3\columnwidth]{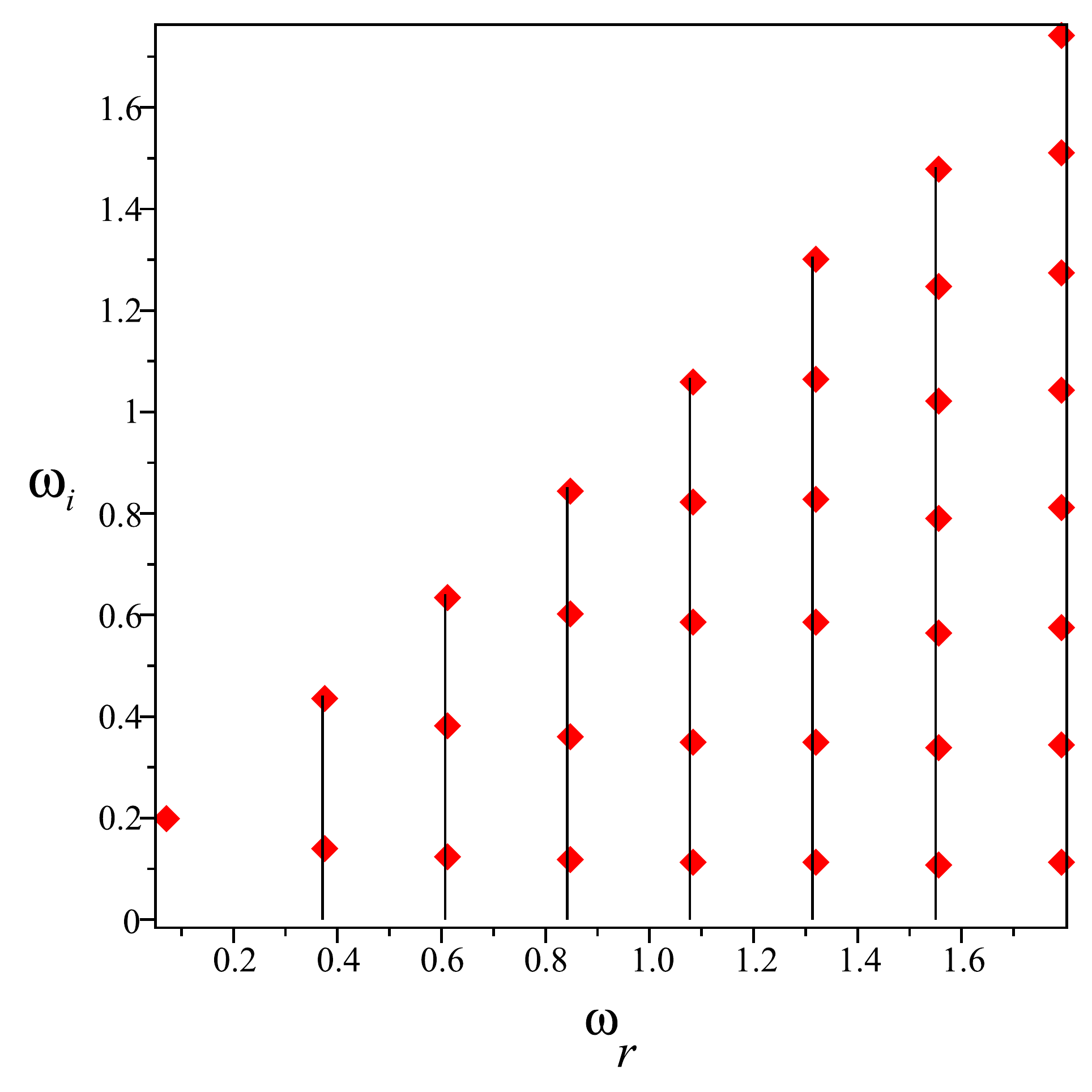}}
\subfigure[$r_{+}=2,M=0.37$]{\includegraphics[width=0.3\columnwidth]{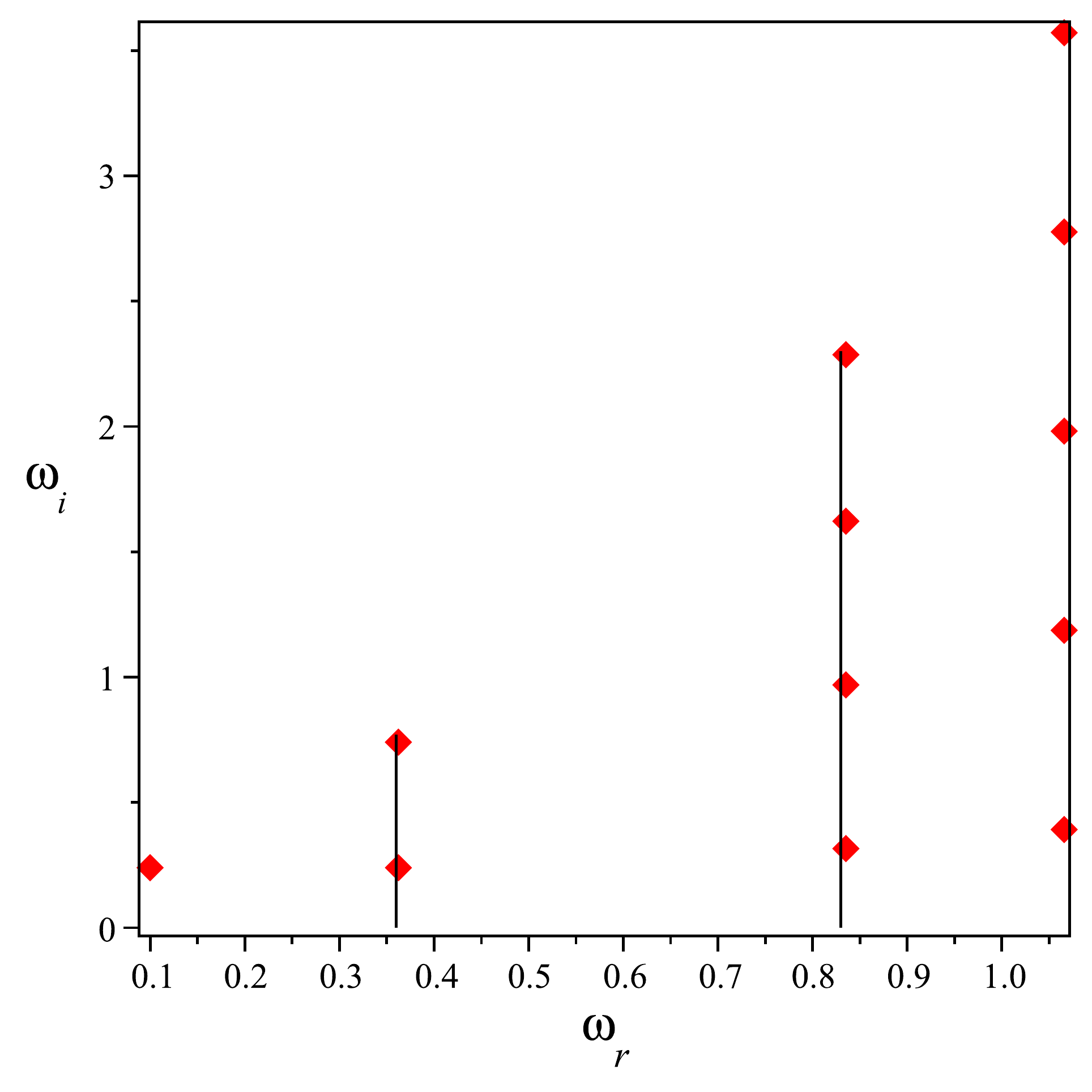}}
\caption{The behavior of $\omega_{i}$ in terms of $\omega_{r}$ for $j=0,1,3,4$ (left) and  $j=0,1,2,3,4,5,6,7$ (right).} 
\label{wwplot}
\end{figure}

In this section, we are interested in calculating the Quasi-Normal Modes (QNMs) of constructed black hole solutions to investigate the dynamical stability of obtained black hole solutions undergoing scalar perturbations. To do so, we used the Klein-Gordon equation for the massless real scalar probe $\phi(r)$ as follows
\begin{equation}
\Box \, \phi(r)=0~.
\end{equation}
{It is important to note that, since we assume that the metric \eqref{metform} is the vacuum solution of the field equation \eqref{eq6} and also the perturbation of the scalar field is second order then can not contribute at the linear perturbation to the field equation \cite{Tsujikawa:2007tg}, \cite{Sebastiani:2010kv}.}
After the separation of variables and changing the radial coordinate to the tortoise coordinate, one can find the second-order Schrodinger-like Ordinary Differential Equation (ODE) for a radial coordinate as follows
\begin{equation}\label{eqqshrod}
\dfrac{d^{2}\varphi(x)}{dx^{2}}+(\omega^{2}-V(x))\varphi(x)=0~,
\end{equation}
where
\begin{equation}
V(r)=h\left(\dfrac{j(j+1)}{r^2}+\dfrac{h^{\prime}}{r}\right),\;\;\;\;\;\;\;x=\int \dfrac{dr}{h}~,
\end{equation}
here, $h(r)$ has been obtained in \eqref{eq17}. The boundary conditions for asymptotically flat black holes are
\begin{equation}\label{qnmbc}
\varphi \sim e^{\pm i\omega x}~~,\;\;\;\;\;\; x\to \pm \infty ~.
\end{equation}
Now, in order to solve the differential equation (\ref{eqqshrod}) with the conditions \eqref{qnmbc}, we used the Mashhoon's Method \cite{vf},\cite{Mashhoon:1985cya}. According to this method the QNM's of a potential barrier are
related to the bound states of the inverted potential. The frequency of waves must be complex, $ \omega = \omega_{r} + i\omega_{i}$. The imaginary and real parts are related to the damping time scale ($\tau_{i} = 1/\omega_{i}$) and oscillation time scale ($\tau_{r} = 1/\omega_{r}$), respectively. 
The proper QNM's using this method can be obtained as
\begin{equation}
\omega_{r}=\pm \left(V_{0}-\dfrac{\eta^2}{4}\right)^{\dfrac{1}{2}},\;\;\;\;\; \omega_{i}=\eta \left(n+\dfrac{1}{2}\right),
\end{equation}
where
\begin{equation}
V_{0}=V(x_{0}),\;\;\;\;\; \eta^{2}=-\dfrac{1}{2V_{0}}\left. \dfrac{d^{2}V}{dx^{2}}\right\vert_{x_{0}}.
\end{equation}
Our results are presented in tables \eqref{table1}, \eqref{table2} and \eqref{table3} and plotted in figures (\ref{wwplot}). A couple of points should be mentioned. First, from the figures and the tables, one can conclude that the quasi-normal frequencies have a positive imaginary part, which shows that both the Schwarzschild and non-Schwarzschild black holes are stable under the scalar perturbations. Second, it is observed for both black holes, with increasing $j$, the values of the real part of the frequencies increase while the imaginary part of the frequencies decreases, which is a sign of more stability of black holes. Also, as can be seen from the figures and tables with increasing $n$, the imaginary part of the frequencies increase which shows that the black hole becomes more unstable.

\begin{table}[H]
\begin{center}
\setlength{\tabcolsep}{0.7cm}
\begin{tabular} { c  c  c c c}
\hline\hline
 $j$&$n$&$\omega_{r}$ &$\omega_{i}$ \\
 \hline\hline
 0&0&0.2272&0.1805 \\
 
1&0&0.6231&0.1504 \\

&1&0.6231&0.4511 \\
  
 2&0&1.02145 &0.1450 \\

&1&1.02145 &0.4302 \\

&2&1.02145 &0.7251 \\
   
3&0&1.4227&0.14338 \\

&1&1.4227 &0.4301 \\

&2&1.4227&0.7169 \\

&3&1.4227&1.00368 \\

\hline\hline
\end{tabular}
\end{center}
\caption{ The values of the quasinormal frequencies in the (small) non-Schwarzschild geometry for the values of parameters $r_{+}=0.5, M=0.55,\alpha=0.5$. \label{table1}}
\end{table}

\begin{table}[H]
\begin{center}
\setlength{\tabcolsep}{0.7cm}
\begin{tabular} { c  c c c c c c c}
\hline\hline
 $j$&$n$&$\omega_{r}$ &$\omega_{i}$&$j$&$n$&$\omega_{r}$&$\omega_{i}$ \\
 \hline\hline
 0&0&0.0659&0.2055 &5&0&1.3133&0.1187\\

1&0&0.3711&0.1471&&1&1.3133&0.3561\\

&1&0.3711&0.4414&&2&1.3133&0.5935\\

2&0&0.6071&0.1283&&3&1.3133&0.8309\\

&1&0.6071&0.3847&&4&1.3133&1.0683\\

&2&0.6071&0.6413&&5&1.3133&1.3057\\

3&0&0.8418&0.1217&6&0&1.5501&0.1140\\

&1&0.8418&0.3651&&1&1.5501&0.3421\\

&2&0.8418&0.6083&&2&1.5501&0.5701\\

&3&0.8418&0.8517&&3&1.5501&0.7981\\

4&0&1.0773&0.1186&&4&1.5501&1.0262\\
 
&1&1.0773&0.3558&&5&1.5501&1.2542\\

&2&1.0773&0.5929&&6&1.5501&1.4823\\
  
 &3&1.0773 &0.83016&7&0&1.7866&0.1165\\

&4&1.0773 &1.067&&1&1.7866&0.3494\\

\hline\hline
\end{tabular}
\end{center}
\caption{ The values of the quasi-normal frequencies in the (large) Schwarzschild geometry for the values of parameters $r_{+}=1.5, M=0.57,\alpha=0.5$. \label{table2}}
\end{table}

\begin{table}[H]
\begin{center}
\setlength{\tabcolsep}{0.7cm}
\begin{tabular} { c  c  c c c}
\hline\hline
 $j$&$n$&$\omega_{r}$ &$\omega_{i}$ \\
 \hline\hline
 0&0&0.097&0.254 \\
 
1&0&0.3595&0.252 \\

&1&0.3595&0.7564 \\
    
3&0&0.8323&0.3279 \\

&1&0.8323&0.984 \\

&2&0.8323&1.638 \\

&3&0.8323&2.296 \\

4&0&1.063&0.398\\

&1&1.063&1.194\\

&2&1.063&1.99\\

&3&1.063&2.79\\

&4&1.063&3.582\\

\hline\hline
\end{tabular}
\end{center}
\caption{ The values of the quasinormal frequencies in the (large) non-Schwarzschild geometry for the values of parameters $r_{+}=2.0, M=0.37,\alpha=0.5$. \label{table3}}
\end{table}

\section{Conclusion}\label{conc}
$f(R)$ gravity is a natural generalization of GR and has been received extensive investigation in past decades.  In particular, the black hole solutions of this gravity are interesting. To do so, in this paper, we obtained analytical approximate black hole solutions in the framework of $R^{2}$ gravity. To achieve this goal, first, we obtained the near horizon and asymptotic solutions and then use these to obtain a complete solution using a continued-fraction expansion. It is important to note that the continued fraction expansion is applicable just from horizon to infinity. Two general types of solutions numerically have been obtained, i.e. black hole solutions and the solutions without horizon and singularity. In addition, the black hole solutions are categorized in two types, i.e. Schwarzschild and non-Schwarzschild-like solutions. Our results showed that there are obvious differences between Schwarzschild and non-Schwarzschild-like solutions. In the non-Schwarzschild-like solutions unlike the Schwarzschild case by increasing the radius of horizon mass and entropy decrease while temperature increases. We studied the stability of the solutions and  we found that the solutions thermodynamically and dynamically are stable. We would like to emphasize that in this work similar to our previous works \cite{Sajadi:2020axg}, \cite{Sajadi:2022ybs}, we assumed that the near horizon constant $h_{1}$ is a function of $r_{+}$. This assumption is eligible, because $h_{1}$ is proportional to temperature according to Equation (\ref{eq20}).

Another interesting direction to extend our work will be to investigate the non-vacuum, rotating black hole, and other solutions of the theory by using the continued-fraction expansion.
\section*{Acknowledgements}
 We would like to thank the referee for his/her fruitful comments which help us to improve the presentation of the manuscript. 
 SNS and SHH thank the support of Iran National Science Foundation 99022223. Also, thanks to Farid Charmchi for useful discussions on numerical programing.

\appendix

\section{Explicit Terms in the Continued-Fraction Expansion}\label{appa}

We present terms up to order 4 in the continued-fraction expansion:
\begin{align}
&\epsilon=-\dfrac{H_{1}}{r_{+}}-1,\,\,\,\, a_{1}=-1-a_{0}+2\epsilon+r_{+}h_{1},\,\,\,\, a_{2}=-{\dfrac {4a_{1}-5\epsilon+1+3 a_{0}+ h_{2}r_{+}^{2}}{{ a_{1}}}}
\nonumber \\
&a_{3}=-\dfrac{1}{{a_{1}}{a_{2}}}[-{h_{3}}{{r_{+}}}^{3}+{a_{1}}{{a_{2}}}^{2}+5{a_{1}}{a_{2}}+6{a_{0}}+10{a_{1}}-9\epsilon+1]  
\label{cfrac-a}\\
&a_{4}=-\dfrac{{h_{4}}{{r_{+}}}^{4}+{a_{1}}{{a_{2}}}^{3}+2{a_{1}}{{a_{2}}}^{2}{a_{3}}+{a_{1}}{a_{2}}{{a_{3}}}^{2}+6{a_{1}}{{a_{2}}}^{2}+6{a_{1}}{a_{2}}{a_{3}}+15{a_{1}}{a_{2}}+10{a_{0}}+20{a_{1}} -14\epsilon+1}{{a_{1}}{a_{2}}{a_{3}}} \nonumber 
\end{align}

\section{Near Horizon Constants}\label{app2}
Here, we present some near horizon constants regarding section \eqref{eq7} as follows:
\begin{align}
h_{5}&=\dfrac{1}{420\alpha^{2}r_{+}^{8}h_{1}^{3}}[-344r_{+}^{3}\alpha^{2}h_{1}h_{2}-3r_{+}^{6}h_{2}
\alpha h_{1}^{2}-372r_{+}^{5}\alpha^{2}h_{1}^{3}h_{2}-556\alpha^{2}r_{+}^{4}h_{2}h_{1}^
{2}+12r_{+}^{6}\alpha^{2}h_{2}^{2}h_{1}^{2}\nonumber\\
&-80r_{+}^{5}\alpha^{2}h_{1}h_
{2}^{2}+28\alpha r_{+}^{4}h_{2}+80\alpha r_{+}^{3}h_{1}-2r_{+}^{4}+2 r_{+}^{5}\alpha h_{1}h_{2}-20 \alpha^{2}r_{+}^{6}h_{2}^{3}+5\alpha r_{+}^{6}h_{2}^{2}-12\alpha r_{+}^{2}\nonumber\\
&+16\alpha^{2}r_{+}^{2}h_{1}^{2}-146 r_{+}^{5}h_{1}^{3}\alpha+1308 \alpha^{2}r_{+}^{4}h_{1}^{4}-2212\alpha^{2}r_{+}
^{3}h_{1}^{3}+100r_{+}^{4}\alpha h_{1}^{2}-60\alpha^{2}r_{+}^{4}h_{2}^{2}+96\,{\alpha}^{2}r_{+}^{2}h_{2}\nonumber\\
&+256{\alpha}^{2}r_{+}h_{1}+2r_{+}^{5}h_{1}-16\,{
\alpha}^{2}
],
\end{align}
and
\begin{align}
h_{6}&=-\dfrac{1}{25200\alpha^{2}r_{+}^{10}h_{1}^{4}}[-106r_{+}^{4}-832\alpha r_{+}^{2}-24r_{+}^{5}h_{1}-1632{\alpha}^{2}+165r_{+}^{6}h_{1}^{2}-42r_{+}^{7}
h_{1}^{3}+3504r_{+}^{3}{\alpha}^{2}h_{1}h_{2}\nonumber\\
&-1374r_{+}^{6}h_{2}\alpha h_{1}^{2}-32112r_{+}^{5}\alpha^{2}h_{1}^{3}h_{2}-33048{\alpha}^{2}r_{+}^{4}h_{2
}h_{1}^{2}-1752r_{+}^{6}{\alpha}^{2}h_{2}^{2}h_{1}
^{2}-14688r_{+}^{5}{\alpha}^{2}h_{1}h_{2}^{2}\nonumber\\
&+3692r_{+}^{5}\alpha h_{1}h_{2}+4848{\alpha}^{2}r_{+}^{3}h_{1}^{3}-158736{\alpha}^{2}r_{+}^{4}h_{1}^{4}+6420
r_{+}^{5}h_{1}^{3}\alpha +3600{\alpha}^{2}r_{+}^{2}h_{1}^{2}-2304{\alpha}^{2}
r_{+}^{6}h_{2}^{3}\nonumber\\
&+1084\alpha r_{+}^{6}h_{2}^{2}+216\alpha r_{+}^{4}h_{2}+1976\,
\alpha r_{+}^{3}h_{1}+4200r_{+}^{4}\alpha h_{1}^{2}+2064{\alpha}^{2}r_{+}^{4}h_{2}^{2}+2352{\alpha}^{2}r_{+}^{2}h_{2}+\nonumber\\
&8160\,{\alpha}^{2}r_{+}h_{1}-97r_{+}^{6}h_{2}+120h_{2}^{3}r_{+}^{8}\alpha -
480 h_{2}^{4}r_{+}^{8}{\alpha}^{2}+90144{\alpha}^{2}r_{+}^{5}h_{1}^{5}-9882r_{+}^{6}h_{1}^{4}\alpha +90r_{+}^{7}h_{2}h_{1}-\nonumber\\
&1680r_{+}^{7}{\alpha}^{2}h_{1}h_{2}^{3}-72h_{2}^{2}r_{+}^{8}\alpha h_{1}^{2}+288h_{2}^{3}r_{+}^{8}{\alpha}^{2}h_{1}^{2}-29664r_{+}^{6}{\alpha}^{2}h_{1}^{4}h_{2}+1008r_{+}^{7}{\alpha}^{2}h_{2}^{2}h_{1}^{
3}+60r_{+}^{7}\alpha h_{1}^{3}h_{2}\nonumber\\
&-180r_{+}^{7}
\alpha h_{2}^{2}h_{1}
].
\end{align}

\section{Misner-Sharp mass}\label{massMS}
Here, we compute the Misner-Sharp mass with equation (4.16) of \cite{Cai:2009qf} as follows:
\begin{equation}
E=\dfrac{r}{2G}\left[(1-h^{a b}\partial_{a}r\partial_{b}r)f_{R}+\dfrac{r^2}{6}(f-Rf_{R})\right],
\end{equation}
where $x^{a}$ is the coordinate on a two-dimensional spacetime with $a=t,r$.
By inserting metric \eqref{metform} into the above definition for mass and using \eqref{eqqfReq}, one can get
\begin{align}\label{eqqap56}
E&=\dfrac{r}{2}(1-h)+\dfrac{\alpha}{12r}[12h h^{\prime\prime\prime}r^3-8r^2h^{\prime\prime}(1-7h+rh^{\prime})-16r^2h^{\prime 2}+8rh^{\prime}(h-4)-r^4h^{\prime\prime 2}\nonumber\\
&+4(1-h)(7h+5)].
\end{align}
Using the asymptotic metric \eqref{eq70}, one can get
\begin{equation}
E=-\dfrac{H_{1}}{2}+\dfrac{5\alpha H_{1}}{r^2}+\alpha^2\left[\dfrac{100H_{1}}{r^4}+\dfrac{80H_{1}^2}{r^5}\right]+\mathcal{O}(\alpha^3)
\end{equation}
If we assume $H_{1}=-2M$, the total energy of black hole becomes
\begin{equation}
E=M-\dfrac{10\alpha M}{r^2}-\alpha^2\left[\dfrac{200M}{r^4}-\dfrac{320M^2}{r^5}\right]+\mathcal{O}(\alpha^3),
\end{equation}
which is the same as ADM mass. 
%On the other hand, the total mass \eqref{eqqap56} in terms of event horizon $r_{+}$ becomes
%\begin{equation}\label{eqqapp57}
%E=\dfrac{r_{+}}{2}-\dfrac{\alpha}{3r_{+}}\left[4r_{+}^{2}f_{2}(r_{+})+8r_{+}f_{1}(r_{+})-5+r_{+}^{4}f_{2}^{2}(r_{+})+4f_{1}^{2}(r_{+})r_{+}^{2}
%+4f_{1}(r_{+})f_{2}(r_{+})r_{+}^3\right]\neq 2TS
%\end{equation}
%As can be seen, it apparently is different from \eqref{eqmass}. But, if we consider \eqref{eqqapp57} as a mass in the first law of thermodynamics \eqref{eqfirstlaw}, and using \eqref{eqq36q}, one can get the mass explicitly in terms of $\alpha$ and $r_{+}$ as follows
%\begin{small}
%\begin{align}
%M_{MS}=\dfrac{r_{+}}{2}-&\dfrac{3}{2}{r_{+}}{\it RootOf} \left( {e^{\it \_Z}}{r_{+}}^{2
%}+8{c_{1}}\alpha-8\alpha{\it Ei} \left(1,-{\it \_Z}\right)\right)-\nonumber\\
%&\dfrac{3}{16}{\frac {{{r_{+}}}^{3} \left( {\it RootOf} \left( {{e}^{{\it \_Z}}}{{r_{+}}}^{2}+8{\it c_{1}}\alpha-8\alpha{\it Ei}\left( 1,-{\it \_Z}\right)\right) \right) ^{2}}{
%\alpha}}.
%\end{align}
%\end{small}
%For small $\alpha\ll 1$, we have
%\begin{small}
%\begin{align}\label{eqqMMS}
%M_{MS}\sim \dfrac{r_{+}}{2}-&\dfrac{3r_{+}}{2}{\it RootOf}\left(c_{1}-{\it Ei}\left(1,-{
%\it \_Z}\right)\right) -\nonumber\\
%&\dfrac{3r_{+}^{3}}{16\alpha}[{\it RootOf}
 %\left(c_{1}-{\it Ei}\left(1,-{\it \_Z}\right)\right)]
 %[{\it RootOf} \left(c_{1}-{\it Ei}\left( 1,-{\it \_Z}
 %\right)\right)+1]+\mathcal{O}(\alpha^{-2}).
%\end{align}
%\end{small}
While the mass \eqref{eqmass} in the case of small $\alpha\ll 1$ becomes 
\begin{align}\label{eqqour}
M&\sim i\sqrt{c_{1}}r_{+}+\dfrac{8c_{1}+3i\sqrt{c_{1}}}{r_{+}}\alpha+\mathcal{O}(\alpha^{2}),\nonumber\\
&\sim - i\sqrt{c_{1}}r_{+}+\dfrac{8c_{1}-3i\sqrt{c_{1}}}{r_{+}}\alpha+\mathcal{O}(\alpha^{2}),\nonumber\\
&\sim -\dfrac{16c_{1}}{r_{+}}\alpha+\dfrac{96c_{1}\alpha^2}{r_{+}^{3}}+\mathcal{O}(\alpha^{3}),
\end{align}
for three solutions of $ h_{1} $ in Eqs. \eqref{eqq36q}-\eqref{eqqab}.  It is straightforward to show that for $c_{1}=-1/4$, the asymptotic value of the mass \eqref{eqqour}  is the same as Schwarzschild mass.\\

\end{document}